\documentclass[twocolumn,showpacs,preprintnumbers,amsmath,amssymb]{revtex4}

\usepackage{graphicx}
\usepackage{dcolumn}
\usepackage{bm}
\usepackage{latexsym}
\usepackage{amsfonts}
\usepackage{amssymb}
\usepackage{amsmath}
\usepackage[usenames]{color}

\begin{document}
\preprint{KUNS-2200}

\title{Chiral Primordial Gravitational Waves from a Lifshitz Point}

\author{Tomohiro Takahashi}
\author{Jiro Soda}

\affiliation{Department of Physics,  Kyoto University, Kyoto, 606-8501, Japan
}

\date{\today}

\begin{abstract}
We study primordial gravitational waves produced during inflation
in quantum gravity at a Lifshitz point proposed by Ho${\rm\check{r}}$ava. 
Assuming power-counting renormalizability, 
foliation preserving diffeomorphism invariance, and the condition of detailed balance,
 we show that primordial gravitational waves are
circularly polarized due to parity violation.
The chirality of primordial gravitational waves 
is a quite robust prediction of quantum gravity at a Lifshitz point
which can be tested through observations of cosmic microwave background radiation
and stochastic gravitational waves. 
\end{abstract}

\maketitle

\section{Introduction}

For a long time, the inflationary scenario has been regarded as 
 an academic theory which provides an elegant solution
 to conceptual problems in cosmology 
 such as the flatness problem, the horizon problem, and 
the origin of structure of the universe. However,
the fact that all of recent cosmological observations strongly support
 the inflationary scenario with high precision encourages us 
 to take the inflation more seriously. Then, taking into account
 that the inflation magnifies microscopic scales to macroscopic ones,
 it is reasonable to regard the inflation as a probe to investigate
  physics at the Planck scale, namely, quantum gravity. 
 
It is widely believed that string theory is a promising candidate for quantum
gravity. However, it is premature to discuss a Planckian regime of the universe
using string theory. Therefore, so far, study of trans-Planckian effect on
inflationary predictions has been
 phenomenological~\cite{Martin:2000xs,Brandenberger:2000wr}. 
In fact, there are various phenomenological models which can mimic trans-Planckian physics
and lead to a modification of the power spectrum of curvature 
perturbations~\cite{Brandenberger:2002sr}.
It is known that these quantitative trans-Planckian corrections 
are suffered from severe constraints due to the 
backreaction problem~\cite{Tanaka:2000jw,Starobinsky:2001kn}.
However, there may be more qualitative effects due to trans-Planckian physics.
For example, polarization of primordial gravitational waves could be
an important smoking gun of trans-Planckian 
physics~\cite{Lue:1998mq,Cai:2007xr,Contaldi:2008yz}. 
In fact, a parity violating gravitational Chern-Simons term which is 
ubiquitous in string theory can generate circular polarization in primordial
 gravitational waves~\cite{Choi:1999zy,Alexander:2004us,Lyth:2005jf}.
  However, it has been shown that the effect of parity violation is negligibly small
   for the slow roll inflation~\cite{Alexander:2004wk}.
Recently, it is argued that  sizable circular polarization
 could be generated~\cite{Satoh:2007gn,Satoh:2008ck}
 by resorting to a peculiar feature due to the Gauss-Bonnet term~\cite{Kawai:1998ab}.
One defect in these models is the appearance of divergence 
in one of circular polarization modes. 
This divergence suggests necessity of a consistent quantum theory of gravity.

Recently, quantum gravity at a Lifshitz point which is power-counting renormalizable 
is proposed by Ho${\rm\check{r}}$ava~\cite{Horava:2008ih,Horava:2009uw}. 
 In contrast to string theory, the theory is not intended to be 
 a unified theory but just quantum gravity in 4-dimensions. 
 In this ``small" framework, 
 one can discuss trans-Planckian effect on cosmology in a self-consistent manner. 
 In Ho${\rm {\check r}}$ava's formulation of quantum gravity, 
the action necessarily contains a Cotton tensor, which violates parity invariance. 
Hence, we can expect circular polarization 
of primordial gravitational waves. 
Moreover, there exists no divergence in this model. 
 Then, the purpose of this paper is to calculate
 degree of circular polarization during inflation 
and show observability of chiral primordial gravitational waves
 which is a robust prediction of quantum gravity at a Lifshitz point.

\section{QG at a Lifshitz point}

The quantum gravity proposed by Ho${\rm {\check r}}$ava can be characterized
by anisotropic scaling at an ultraviolet fixed point
$
	{\bf x}\rightarrow b{\bf x},\ t\rightarrow b^3t , 
$
where $b$, ${\bf x}$ and $t$ are a scaling factor, 
spacial coordinates and a time coordinate, respectively.
This scaling guarantees the renormalizability of the theory~\cite{Horava:2009uw}.
Because of the anisotropic scaling, the time direction plays a privileged role. 
In other words, the spacetime has a codimension-one foliation structure in which 
leaves of the foliation are hypersurfaces of constant time. 
Since the spacetime has the anisotropic scaling and the foliation structure, 
 the theory is not diffeomorphism invariant but invariant under
the foliation-preserving diffeomorphism defined by 
$
	{\tilde x}^{i}={\tilde x}^{i}(x^j,t),\ {\tilde t}={\tilde t}(t). 
$
 Here, indices $i,j,k,\cdots$ represent spacial coordinates.
 
To describe the foliation, it is convenient to use ADM decomposition of the metric
$ds^2 = -N^2 dt^2 + g_{ij} (dx^i +N^i dt)(dx^j + N^j dt)$, where
 $N$, $N_i$, and $g_{ij}$ are the lapse function, the shift function and 
 the 3-dimensional induced metric, respectively.  
In order for the theory to be unitary,  
the number of time derivative should be at most two in the action. 
The renormalizable kinetic part is then given by
\begin{eqnarray}
S_K =  \frac{2}{\kappa^2} \int dt d^3x \sqrt{g} N 
   \left( K^{ij}K_{ij}-\lambda K^2 \right)   \ , 
   \label{kinetic}
\end{eqnarray}
where $K_{ij}$ is the extrinsic curvature of the constant time hypersurface defined by
$
	K_{ij}= \left(\dot{g}_{ij}-N_{i|j}-N_{j|i}\right) /2N ,
$
and $K$ is the trace part of $K_{ij}$. 
Note that $\kappa$ and $\lambda$ are  dimensionless
coupling constants which run according to the renormalization group flow.
Hereafter, we assume $\lambda$ has already settled down to the infrared
fixed point $\lambda =1$ at the beginning of the inflation
although we keep $\lambda$ in subsequent formulas.

The most crucial assumption of Ho${\rm {\check r}}$ava's theory is
the detailed balance condition
\begin{eqnarray}
  S_V = \frac{\kappa^2}{8} \int dt d^3 {\bf x} \sqrt{g} N E^{ij}
  {\cal G}_{ijkl} E^{kl} \ ,
\end{eqnarray}
where we have defined 
$
   \sqrt{g} E^{ij} = \delta W[g_{ij}] / \delta g_{ij}
$
with some functional $W$.   
Here, we introduced the inverse of De Witt metric
$
  {\cal G}^{ijkl} = ( g^{ik} g^{jl} + g^{il} g^{jk} )/2 
                     - \lambda g^{ij} g^{kl}   . 
$
The renormalizability of the theory requires $E^{ij}$ must be third order
in spatial derivatives. The requirement uniquely selects the Cotton tensor
\begin{eqnarray}
C^{ij}=\varepsilon^{ikl}\nabla_k\left(R^{j}_{l}-\frac{1}{4}R\delta^{j}_{l}\right) \ ,
\label{cotton}
\end{eqnarray}
where $\epsilon^{ijk}$ denotes the totally antisymmetric tensor
and $R_{ij}$ and $R$ are the 3-dimensional Ricci tensor and
 Ricci scalar.
Including relevant deformations, we have
\begin{eqnarray}
   W &=& \frac{1}{w^2} \int d^3 {\bf x} \sqrt{g}  \epsilon^{ijk}
   \left( \Gamma^m_{il} \partial_j \Gamma^l_{km}
   + \frac{2}{3} \Gamma^n_{il}\Gamma^l_{jm}\Gamma^m_{kn} \right) \nonumber  \\
   &&   + \mu\int d^3 {\bf x} \sqrt{g} \left(  R -2\Lambda_w \right)  \ ,
\end{eqnarray}
where $\Gamma^i_{jk}$  and $\Lambda_w$ are Christoffel symbols 
 and the 3-dimensional ``cosmological constant", respectively.
Here, we have introduced new coupling constants $w$ and $\mu$.
Note that the first two terms lead to the Cotton tensor.
Thus, we obtain the potential part of
 the 4-dimensional action~\cite{Horava:2009uw}
\begin{eqnarray}
	S_V &=&\int dt d^3x \sqrt{g}  N \nonumber\\
	 && \times 
  \Bigl[ -\frac{\kappa^2}{2 w^4}C^{ij}C_{ij}
	 +\frac{\kappa^2\mu}{2 w^2}\varepsilon^{ijk}R_{il}R^{l}_{k|j}
       -\frac{\kappa^2\mu^2}{8}R_{ij}R^{ij}\nonumber\\
	 &\ &\hspace{0.3cm}+\frac{\kappa^2\mu^2}{8(1-3\lambda)}\left(\frac{1-4\lambda}{4}R^2+\Lambda_wR-3\Lambda_w^2\right)   \Bigr] \ , 
       \label{action}
\end{eqnarray}
where a stroke $|$ denotes a covariant derivative with respect to spacial coordinates. 
In the above action (\ref{action}), the coefficient of scalar curvature $R$
 is $\kappa^2\mu^2\Lambda_w/8(1-3\lambda)$, 
 then the gravitational constant become negative in the low energy limit 
unless $\Lambda_w/(1-3\lambda)>0$.

In addition to this gravity sector, we consider the action for an inflaton $\phi$
\begin{eqnarray}
 S_M = \int dt d^3 {\bf x} \sqrt{g} N
  \left[- \frac{1}{2} \partial^\mu \phi \partial_\mu \phi
                 - V(\phi)  \right] \ .
     \label{inflaton}
\end{eqnarray}
Let us assume the slow roll inflation and take the slow roll limit.
Then, we can replace the action (\ref{inflaton}) with
 the effective cosmological constant $\bar{\Lambda}$, namely, we
have $S_M = - \int dt d^3 {\bf x} \sqrt{g} N \bar{\Lambda}$. 

 Thus, the total action is given by $S= S_K +S_V +S_M$.
  The total action $S$ breaks the detailed balance condition
   softly~\cite{Horava:2009uw}. 
  It should be emphasized that the total action $S$ reduces to
  the conventional Einstein theory at low energy.

\section{Primordial gravitational waves}

Let us consider the background spacetime with spatial isotropy and homogeneity
$
	ds^2= -dt^2+a(t)^2\delta_{i j}dx^idx^j ,
$
where $a$ is the scale factor.
Using this metric ansatz, we can get the  Friedmann equation with $\lambda=1$
\begin{eqnarray}
	\frac{\dot{a}^2}{a^2}=\frac{\kappa^2}{12}\left(\bar{\Lambda}-\frac{3\kappa^2\mu^2\Lambda_w^2}{16}\right)\equiv H^2 \ .
\end{eqnarray}
 The above equation leads to de Sitter spacetime, 
$
	a(t) \propto  e^{Ht} \label{solution}
$.
 Here, we assumed  $\bar{\Lambda}> 3\kappa^2\mu^2\Lambda_w^2 /16 $. 
Note that if we chose $\bar{\Lambda}=0$, there is no Minkowski solution. 
Hence, there must exist residual vacuum energy in the matter sector at the end of the day.
This is related to the issue of the cosmological constant, which is
beyond the scope of this paper.

Now, we consider  tensor perturbations 
$
	ds^2=-dt^2+a(t)^2(\delta_{ij}+h_{ij}(t,{\bf x}))dx^idx^j  , 
      \label{perturbation}
$
where $h_{ij}$ satisfies the transverse-traceless conditions. 
Substituting this metric into the total action, we  
obtain the quadratic action
\begin{eqnarray}
	\delta^2 S&=&\int dtd^3x a^3\Bigl[\frac{1}{2\kappa^2}\dot{h}^{i}_{j}\dot{h}^{j}_{i}+\frac{\kappa^2}{8 w^4a^6}\Delta^2 h^{i}_{j}\Delta h^{j}_{i}\nonumber\\
	          &\ &\  +\frac{\kappa^2\mu}{8 w^2a^5}\epsilon^{ijk}
                \Delta h_{il}\Delta h^{l}_{k|j}-\frac{\kappa^2\mu^2}{32a^4}
                \Delta h^{i}_{j}\Delta h^{j}_{i}\nonumber \\
		    &\ &\hspace{2.5cm} 
          +\frac{\kappa^2\mu^2\Lambda_w}{32(1-3\lambda)a^2}h^{i}_{j}\Delta h^{j}_{i}
	\Bigr], \label{qaction}
\end{eqnarray}
where $\Delta$ represents the Laplace operator. 
The transverse traceless tensor $h_{ij}$ can be expanded in terms of plane waves
with wavenumber ${\bf k}$ as
\begin{eqnarray}
	h_{ij}(t,{\bf x})=\sum_{A=R,L}
       \int\frac{d^3{\bf k}}{(2\pi)^3}\psi^{A}_{{\bf k}}(t)
       e^{i{\bf k}\cdot{\bf x}} p_{ij}^{A} \ ,
      \label{expansion}
\end{eqnarray}
where $p_{ij}^{A}$ are circular polarization tensors which are defined by
$
	 ik_{s}\epsilon^{rsj}p_{ij}^{A} = k\rho^{A}p^{r}_{\ i}{}^{A}  
$ 
\cite{Satoh:2007gn}. 
Here, $\rho^{R} =1$ and $\rho^{L} =-1$ modes are called the right handed mode and
the left handed mode, respectively. 
We also impose normalization conditions
$
	p^{*}{}^{i\ A}_{\ j}p^{j\ B}_{\ i} = \delta^{AB}  ,
$
where $p^{*}{}^{i\ A}_{\ j}$ is the complex conjugate of $p{}^{i\ A}_{\ j}$.
Substituting the expansion (\ref{expansion}) into the gravitational action (\ref{qaction}),
we obtain
\begin{eqnarray}
	 \delta^2 S&=&\sum_{A=R,L}\int dt\frac{d^3{\bf k}}{(2\pi)^3} a^3
       \Bigl[\frac{1}{2\kappa^2}|\dot{\psi}^{A}_{\bf k}|^2\nonumber\\
			&\ &\ -\Bigl\{\frac{\kappa^2 k^6}{8 w^4 a^6}
         -\rho^{A}\frac{\kappa^2\mu k^5}{8 w^2 a^5}
         +\frac{\kappa^2 \mu^2 k^4}{32a^4}\nonumber \\
	    &\ &\hspace{2.5cm} +\frac{\kappa^2\mu^2\Lambda_w k^2}{32(1-3\lambda)a^2}
	\Bigr\}|\psi^{A}_{\bf{k}}|^2\Bigr]  \ . 
\end{eqnarray}
Using the variable $v^{A}_{\bf k}\equiv a\psi^{A}_{\bf k}$
 and conformal time $\eta$ defined by $d\eta/dt=1/a$, 
 we obtain the equations of motion 
\begin{eqnarray}
	\frac{\partial^2}{\partial \eta^2}v^{A}_{\bf{k}}
      +\left(  k^{A\ 2}_{eff}
      -\frac{2}{\eta^2}\right)v^{A}_{\bf{k}}=0 \ ,
       \label{vEOM}
\end{eqnarray}
where we used $a=- 1/H\eta$ and defined
$
 k^{A\ 2}_{eff} = \alpha^2 k^2 \left\{1+
	\beta (\alpha k \eta )^2
      (1+\rho^{A}\gamma \alpha k \eta )^2\right\}  .
$
We have also  defined
\begin{eqnarray}
	\alpha^2=\frac{\kappa^4\mu^2\Lambda_w}{16(1-3\lambda)} \ , \  
      \beta=H^2\frac{1-3\lambda}{\Lambda_w \alpha^2} \ ,  \  
      \gamma=H\frac{2}{ w^2 \mu \alpha} \ .
\end{eqnarray}
Here, $\alpha$ is "the emergent speed of light"~\cite{Horava:2009uw}
 and $\beta$ and $\gamma$ are dimensionless parameters.

Since there appears $\rho^{A}$ in Eq.(\ref{vEOM}), 
the evolution of right handed mode is different from that of left handed mode. 
Hence, the dimensionless parameter $\gamma$ characterizes
 "the parity violation".
If $\beta=0$  and $\alpha$ is exactly the speed of light,
 Eq.(\ref{vEOM}) becomes the equation for gravitational waves
  in pure de Sitter background in Einstein theory. Then 
$\beta$ measures the "deviation from Einstein theory".  
 
\section{circular polarization}

Now, we calculate the power spectrum $|\psi_{{\bf k}}^{A}|^2$ numerically
and evaluate degree of circular polarization of primordial gravitational waves.

\begin{figure}[htbp]
 \begin{center}
  \includegraphics[width=80mm]{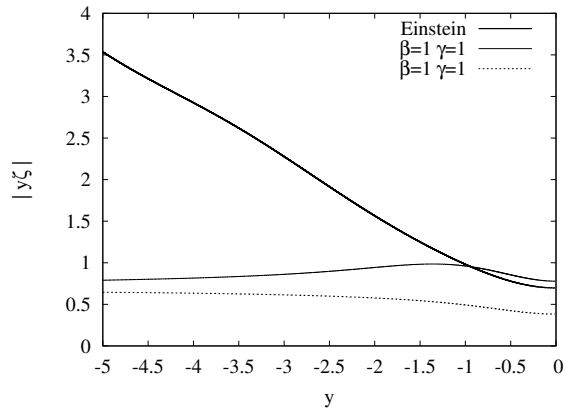}
 \end{center}
 \caption{The time evolution of the power is depicted. The thick solid line represents 
 the evolution for the conventional Einstein gravity.
 The thin solid line and the dotted line show the time evolution of
  right handed mode and left handed mode, respectively. }
 \label{fig:1}
\end{figure}

For the numerical analysis, it is convenient to introduce dimensionless
 variables $k^{'}\equiv \alpha k / H$ and $y\equiv k^{'}H\eta$.  
Using these variables and the transformation  
$\zeta^{A}\equiv \sqrt{k^{'}H}v_{\bf k}^{A}/\kappa$, 
we can write down the basic equation
\begin{eqnarray}
	\frac{d^2}{dy^2}\zeta^{A}+ \omega^2(y)\zeta^{A}=0 \ ,
      \label{ODE}	
\end{eqnarray}
where
$
	\omega^2(y)=1+\beta y^2(1+\rho^A\gamma y)^2- 2/y^2  . 
      \label{omega}
$
Since WKB approximation is pretty good in the asymptotic past $y\rightarrow -\infty$, 
we can choose the adiabatic vacuum as the initial condition. 
More precisely, we set the positive frequency modes as
\begin{eqnarray}
 \zeta^{A}=\frac{1}{\sqrt{2 \omega (y)}}
 \exp\left\{-i\int_{y_i}^{y} \omega (y^{'})dy^{'}\right\} \ . 
 \label{barIC}
\end{eqnarray}
On superhorizon scales $y\rightarrow 0$,
 Eq.(\ref{ODE}) has asymptotic solution $\zeta^{A}=C^{A}/y+D^{A}y^2$
 with constants of integration $C^A$ and $D^A$. 
 Hence, the power spectrum defined by
$
	k^3 |\psi_{\bf k}^{A}|^2= k^3 \left| v_{\bf k}^{A} / a \right|^2
      = \kappa^2 H^2 \left|y\zeta^{A}\right|^2 / \alpha^3  
$
reduces to
\begin{eqnarray}
	 k^3 |\psi_{\bf k}^{A}|^2
       =\frac{\kappa^2 H^2}{\alpha^3} \left|C^{A}\right|^2\label{spec}
\end{eqnarray}
on superhorizon scales $y\rightarrow 0$.
 So, we only need to calculate $C^{A}$ using Eq.(\ref{ODE}) with the
initial condition (\ref{barIC}). Notice that the power spectrum 
$P(k) = k^3 |\psi_{\bf k}^{A}|^2$ is scale free.
From the mode function (\ref{barIC}), we see the vacuum depends on the chirality. 
In the WKB regime, the amplitude of the right handed mode grows, while that of 
the left handed mode decays. These two effects make the difference.
In Fig.\ref{fig:1},  we plotted the time evolution of the power
 for a right handed mode and a left handed mode and also displayed 
 the case of Einstein theory for comparison.
 Clearly, one can see that the differences of initial amplitude
 and the growth rate during WKB regime lead to circular polarization. 

\begin{figure}[htbp]
 \begin{center}
  \includegraphics[width=80mm]{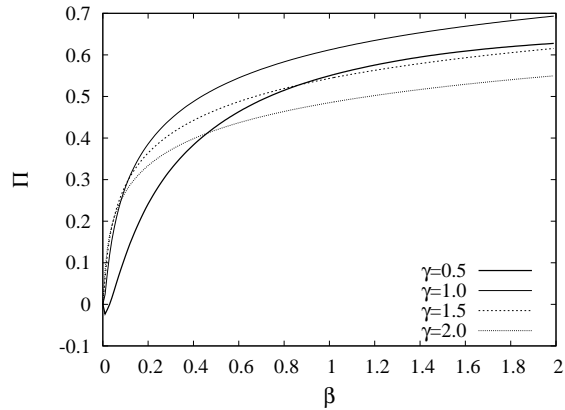}
 \end{center}
 \caption{The degree of circular polarization $\Pi$ for various $\gamma$ 
 as a function of $\beta$ are shown. As can be seen from the figure,
 $\Pi$ grows as the $\beta$ becomes large. The dependence on $\gamma$ is not monotonic
 rather there is a value which gives the maximum polarization for fixed $\beta$.}
 \label{fig:2}
\end{figure}

Now, we are in a position to discuss observability of circular polarization. 
 For this aim, we need to quantify polarization by
  defining degree of circular polarization
\begin{eqnarray}
\Pi = 
 \frac{|\psi_{\bf k}^{R}|^2-|\psi_{\bf k}^{L}|^2}{|\psi_{\bf k}^{R}|^2+|\psi_{\bf k}^{L}|^2}
  = \frac{|C^{R}|^2-|C^{L}|^2}{|C^{R}|^2+|C^{L}|^2}  \ . \label{pi}
\end{eqnarray}
Numerical results are plotted in Fig.\ref{fig:2}.  
There are two possible channels to observe circular polarization
of primordial gravitational waves.
One is the in-direct detection of circular polarization through the
cosmic microwave background radiation,
the required degree of circular polarization has been obtained
as $|\Pi| \gtrsim 0.35 (r/0.05)^{-0.6}$ in \cite{Saito:2007kt}, 
where $r$ is the tensor-to-scalar ratio. 
The relevant frequency of gravitational waves 
in this case is around $f \sim 10^{-17}$ Hz. 
From Fig.\ref{fig:2}, supposing $r=0.05$
and $\gamma=1$, we see we can detect the circular polarization through the temperature
and B-mode polarization correlation if $\beta > 0.2$.
The other is the direct detection of circular polarization,
the required degree of circular polarization has been estimated 
as $\Pi \sim 0.08 (\Omega_{\rm GW} /10^{-15})^{-1} ({\rm SNR}/5)$
around the frequency $f \sim 1$ Hz~\cite{Seto:2006hf}, where
$\Omega_{\rm GW}$ is the density parameter of the stochastic gravitational waves
and SNR is the signal to the noise
 ratio~\cite{Seto:2006hf,Seto:2006dz,Seto:2007tn}. 
Here, 10 years observational time is assumed. Taking look at Fig.\ref{fig:2},
one can see it is easy to get the circular polarization of the order of $0.08$
in the present model. Hence, we can prove or disprove quantum gravity 
at a Lifshitz point by these observations.

\section{Conclusion}

We have considered the inflationary scenario in the context of
quantum gravity at a Lifshitz point which is supposed to be a
power-counting renormalizable theory. Because of the detailed balance condition,
the action necessarily contains Cotton tensor which violates
the parity invariance. 
We have calculated degree of circular polarization of
primordial gravitational waves. As a consequence, we find that
chiral primordial gravitational waves exist for generic parameters.
It should be emphasized that the existence of circular polarization is
a robust prediction of the theory. 

In the usual discussions on the trans-Planckian effects, phenomenological approaches
have been adopted and mostly a modification of the spectrum has been discussed.
While, we have used a candidate of quantum gravity and
discussed chirality of primordial gravitational waves.
The point is that we have found a modification of nature of gravitational waves
 rather than a modification of shape of the spectrum for gravitational waves.
It is also apparent that the power spectrum for curvature perturbations 
 is scale free. Thus, quantum gravity at a Lifshitz point 
is consistent with all of current observations. 
Moreover, we have a testable smoking gun of quantum gravity.

There are many issues to be pursued. 
One of those is to find exact solutions which represent black holes. 
 When black hole solutions are found, it is very interesting to examine
 their spacetime structures, thermodynamics and Hawking radiation. 
 It would be also intriguing to apply the idea of the anisotropic 
 scaling in gravity to braneworld cosmology~\cite{Soda:2006wr}. 
 Furthermore, the possibility to generalize the anisotropic scaling in gravity
 to the cases where spatial isotropy is broken would be interesting
 from the point of anisotropic inflationary
  scenarios~\cite{Kanno:2008gn,Watanabe:2009ct}.

After submitting our paper, we found a related work in the archive
where inflation caused by a Lifishitz scalar is considered~\cite{Calcagni:2009ar}.

\begin{acknowledgements}
JS is supported by the Japan-U.K. Research Cooperative Program and 
Grant-in-Aid for  Scientific Research Fund of the Ministry of 
Education, Science and Culture of Japan No.18540262. 
\end{acknowledgements}

\end{document}